
\magnification=1200
\parindent 1.0truecm
\baselineskip=16pt
\hsize=6.0truein
\vsize=8.5truein
\rm
\null


\footline={\hfil}
\vglue 0.8truecm
\rightline{\tt nucl-th/yymmxxx}
\rightline{\bf DFUPG-05-PG-02}
\rightline{\sl (last revised February 2006)}
\vglue 2.0truecm

\centerline{\bf Low--Energy Kaon-Nucleus Interactions at a $\phi$--Factory }
\vglue 1.0truecm

\centerline{\sl P.M. Gensini$^{\dag\ddag}$, G. Pancheri$^{\S}$, 
Y.N. Srivastava$^{\dag}$ and G. Violini$^{\star}$ }
\vglue 0.3truecm

\centerline{\sl $^{\dag}$) Dip. di Fisica dell'Universit\`a di Perugia, Italy,}
\centerline{\sl and I.N.F.N., Sezione di Perugia, Italy.}
\centerline{\sl $^{\S}$) I.N.F.N., Laboratori Nazionali di Frascati, Italy.}
\centerline{\sl $^{\star}$) Dip. di Fisica dell'Universit\`a della Calabria, 
Arcavacata di Rende (Cosenza), Italy,}
\centerline{\sl and I.N.F.N., Laboratori Nazionali di Frascati, Gruppo 
Collegato di Cosenza, Italy.}
\centerline{ $^{\ddag}$) Speaker at the meeting.}
\vglue 1.0truecm

\centerline{\bf ABSTRACT}

Scattering and formation experiments can be performed not only with particle 
beams, but also, as of old, with particle sources. DA$\Phi$NE at the c.m. energy 
of the $\phi$ is a relatively clean source of low--momentum charged and neutral 
kaons. As such, it can allow experiments unthinkable of at conventional kaon 
beams. This talk is dedicated to a presentation both of this viewpoint and of the 
physics that could be learned {\sl only} in this way.
\vglue 1.0truecm

\centerline{Talk presented at}
\centerline{\sl ``Meeting on $e^+ e^-$ Physics Perspectives (non-K Physics)''}
\centerline{\sl L.N.F., Frascati, January 19 -- 20, 2006}

\pageno=0
\vfill
\eject


\footline={\hss\tenrm\folio\hss}

\vglue 1.0truecm
\centerline{\bf Low--Energy Kaon--Nucleus Interactions at a $\phi$--Factory }
\vglue 0.5truecm
\centerline{\sl P.M. Gensini, G. Pancheri, Y.N. Srivastava and G. Violini }

\vglue 1.0truecm
\noindent{\bf 1. Introduction.}
\vglue 0.3truecm

We intend to illustrate in this proposal the possibilities on strangeness 
$-1$, baryonic physics opened by the $\phi$--factory DA$\Phi$NE$^1$, and in 
particular by its detector KLOE$^2$, considered as a huge (by design), active, 
gaseous $^4$He target.

The interest in this field is of a systematic rather than exploratory nature: 
information on low--energy kaon--nucleon interactions is scarce and of a poor 
statistical quality, when compared to the corresponding pion--nucleon ones. As 
an example, just take a look at the two pages dedicated by the PDG booklet$^3$ 
to $K^\pm p$ and $K^\pm d$ total and elastic cross sections: other data do not 
present a rosier perspective$^4$.

The low quality of low--momentum elastic and inelastic scattering data 
reflects in turn on our knowledge of the ``elementary'' parameters of the 
$K N$ interaction, remarkably poorer than in the $SU(3)_f$--related $\pi N$ 
case$^4$. On top of this sorry situation, one must add the problem of fitting 
into the picture the kaonic hydrogen (and deuterium) level--shifts and 
widths, whose recent experimental determinations, despite having finally 
come out after many years with the sign (almost) every theorist expected$^5$, are 
still awaiting an adequate explanation for their magnitude(s)$^6$.

Data at very low momenta and at rest are essential to clarify many of the 
above--mentioned problems$^4$; however, experiments of this kind pose 
formidable problems at conventional fixed--targed machines, some of which can 
be circumvented at a $\phi$--factory. For instance, at the KAON factory that 
was planned for TRIUMF$^7$, beams in the lowest momentum range (from 400 to 
800 $MeV/c$) would have had intensities of $10^6$ -- $10^8$ $K^- s^{-1}$, with 
$K^+$ beams about twice more intense. Already the purity of these beams is 
limited by $K^\pm$ decays in flight: to experiment at momenta below 400 
$MeV/c$ one has to use moderators, which at the same time decrease the kaon 
intensity, degrade the beam resolution, and increase enormously the beam 
contamination at the final target. All these effects make the experiments much 
more complex, overturning all the advantages offered by the higher initial 
beam intensities.

\vglue 1.0truecm
\noindent{\bf 2. Kaon--Nucleus Interactions at DA$\Phi$NE. }
\vglue 0.3truecm

DA$\Phi$NE is the $\phi$--factory (the acronym stands for ``Double Annular 
$\Phi$--factory for Nice Experiments'') of the I.N.F.N. National Laboratories 
in Frascati. From its expected commissioning luminosity$^8$ of $5 \times 
10^{32}\ cm^{-2} s^{-1}$, and an annihilation cross section of about 5 $\mu b$ 
at the $\phi$--resonance peak, one can see that its {\sl two} interaction 
regions should have been the sources of $\simeq 1.2 \times 10^3$ $K^\pm 
s^{-1}$, at a central momentum of 126.9 $MeV/c$, with the momentum resolution 
of $\simeq 1.1 \times 10^{-2}$ due to the very small energy spreads in the 
beams, as well as of $\simeq$ 850 $K_L s^{-1}$, at a central momentum of 110.1 
$MeV/c$, with the slightly poorer resolution of $\simeq 1.5 \times 10^{-2}$.

Both $\pi^\pm$'s and leptons coming out of the two sources are 
easy--to--control backgrounds: the first because the $\pi^\pm$'s, though 
produced at a rate of about 380 $\pi^\pm s^{-1}$ (not counting those from 
$K_S$ decays), come {\sl almost all} from events with three or more final 
particles and can thus be suppressed by momentum and collinearity cuts; the 
second, as well as collinear pions from $e^+ e^- \to \pi^+ \pi^-$, produced at 
much lower rates of order 0.75 $s^{-1}$ (the leptons) or 0.25 $s^{-1}$ (the 
pions), are eliminated by a momentum cut, having momenta about four times 
those of the $K^\pm$'s.

The {\sl two} interaction regions are therefore {\sl small--sized} sources of 
{\sl low--momen-tum, tagged} $K^\pm$'s and $K_L$'s, with {\sl negligible 
contaminations} (after suitable cuts on angles and momenta of the particles 
are applied {\sl event by event}), in an environment of very low background 
radioactivity: this situation is simply unattainable with {\sl conventional} 
technologies at fixed--target machines$^9$, where the impossibility of placing 
experiments too close to the production target limits from below the 
charged--kaon momenta, and kaon decays in flight always contaminate the beams: 
low--momentum experiments are thus possible only with the use of moderators, 
with a huge beam contamination at the target, as well as a large 
final--momentum spread due to straggling phenomena.

It is therefore of interest to consider the feasibility of low--energy, 
$K^\pm N$ and $K_L N$ experiments at DA$\Phi$NE, with respect to equivalent 
projects at machines such as, {\sl e.g.}, the sadly aborted KAON at 
TRIUMF$^{10}$ or the equally ill--fated EHF project$^{11}$.

We shall here try and give an evaluation of rates to be expected in a 
simple--geometry apparatus at DA$\Phi$NE, such as KLOE$^2$. We shall assume 
cylindrical symmetry, with a toroidal target fiducial volume, limited by radii 
$a$ and $a + d$ and of length $l$ (inside and outside of which one can imagine 
a tracking system, surrounded on the outside by a photon detecting system -- 
{\sl e.g} lead--Sci--Fi sandwiches -- and a superconducting, solenoidal coil 
to provide the {\sl moderate} magnetic field {\bf B} needed for momentum 
measurements), filled with a gas at close to atmospheric pressure. For the 
already existing detector KLOE$^2$, we have the additional benefit that the 
gas chamber is fully wired, providing thus a tracking of the charged particles 
akin to the one available in the bubble chambers of old, but without the 
nuisance of the huge dead times of these latter.

One must convert the usual, fixed--target expression for reaction rates to a 
spherical geometry, and also include kaon decays in flight, getting (for 
{\bf B} = 0 or $K^0_L$'s: the general case can be easily treated with slight 
modifications)
$$
d N_r = [ {1 \over \rho^2} ({3 \over {8 \pi}})\ (L \sigma_\phi B_\phi)
\sin^2 \theta e^{- \rho / \lambda} ] \sigma_r \rho_t (\rho^2 d \rho \sin
\theta d \theta d \phi)\ , \eqno (1)
$$
with $\rho$, $\theta$ and $\phi$ spherical coordinates (with the $z$--axis 
oriented along the beam direction), $L$ the machine luminosity, $\sigma_\phi$ 
the annihilation cross section at the $\phi$--resonance peak, $B_\phi$ the 
$\phi$ branching ratio into the desired mode (either $K^+ K^-$ or $K_L K_S$), 
$\sigma_r$ the reaction cross section for the process considered, $\rho_t$ the 
target {\sl nuclear} density, and $\lambda = p_K \tau_K / m_K$ the decay 
length (respectively of 0.954 $m$ for $K^\pm$'s and of 3.429 $m$ for $K_L$'s) 
at the $\phi$--resonance momenta.

Integrating over the fiducial volume, the reaction rate can be cast into the 
simple formula
$$
N_r = {3 \pi \over 4} r d (L \sigma_\phi B_\phi) \rho_t \sigma_r \ , \eqno (2)
$$
with both geometrical acceptance and kaon decay in flight thrown into the 
``reduction factor'' $r$, which we have estimated to take the values 0.50 for 
$K^\pm$'s and 0.72 for $K_L$'s for a fiducial volume defined by $a =$ 10 $cm$, 
$d =$ 50 $cm$ and $L =$ 1 $m$, to represent a ``person--sized'' detector, 
fitting in DA$\Phi$NE's interaction regions. The factors would be closer to 
unity for a detector the size of KLOE: the parameter most influential on $r$ 
is of course $a$ as long as $l$ is at least comparable with the decay lengths 
$\lambda$, due to the angular distribution of the produced kaons; besides, for 
$K^{\pm}$ $r$ increases almost linearly but slowly with increasing field $B$, 
due to the interplay of the increased path length inside the fiducial volume 
on one side, and of the particle decays on the other.

Due to the rather sorry state of our experimental informations in the energy 
region relevant for DA$\Phi$NE, where only data on hydrogen are available, we 
are able to give reliable estimates for the rates only for H$_2$: from the 
{\sl best available} phenomenological analysis (still that of J.K. Kim, dated 
1966--1967$^{12}$) one can {\sl roughly estimate} the corrections to be 
applied (in the impulse approximation only) to other, light targets such as 
D$_2$, $^4$He or $^3$He.

This gives, for a target volume filled by an almost ideal gas at room 
temperature (such as $^4$He/$^3$He or H$_2$/D$_2$), the rates for 
$K^\pm$--initiated processes
$$
N_r = p(\hbox{atm}) \times \sigma_r(\hbox{mb}) \times (4.0\times 10^4\
\hbox{events/y}) \ , \eqno (3)
$$
for a ``Snowmass year'' of $10^7$ s (for $K_L$'s the figure in eq. (3) is 
about the same, because of an approximate compensation between the variations 
in $r$ and $B_\phi$), or, with rough estimates of the partial $K^- p$ cross 
sections at the $\phi$--decay momenta, to about $10^7$ two--body events per 
year in $H_2$ gas at atmospheric pressure, of which about $3.6 \times 10^6$ 
elastic scattering events, $2.4 \times 10^6$ $\pi^+ \Sigma^-$ and about $10^6$ 
for each of the remaining four two--body channels $\pi^0 \Sigma^0$, $\pi^0 
\Lambda$, $\bar K^0 n$, and $\pi^- \Sigma^+$. The above rates are enough to 
measure angular distributions in all channels, and also the polarizations for 
the self--analyzing final--hyperon states, particularly for the decays 
$\Lambda \to \pi^- p$, $\pi^0 n$ (asymmetry $\alpha \simeq 0.64$) and 
$\Sigma^+ \to \pi^0 p$ ($\alpha \simeq -0.98$). One could also expect a total 
of about $10^4$ radiative--capture events, which should allow a good 
measurement on the {\sl absolute} rates for these processes as well.

Such an apparatus will need: tracking for incoming and outgoing charged 
particles, time--of--flight measurements (for charged--particle 
identification), a moderate magnetic field (due to the low momenta involved) 
for momentum measurements, and a system of converters plus scintillators for 
photon detection and subsequent geometrical reconstruction of $\pi^0$ and 
$\Sigma^0$ decays, amounting thus to a rather simple (on today's 
particle--physics scale), not too costly apparatus. Mentioning costs, we wish 
to point out that DA$\Phi$NE, though giving the experimenters a very small 
momentum range, saves them the cost of the {\sl separate} tagging system 
needed to reject contaminations in a {\sl conventional} low--energy, 
fixed--target experiment$^9$.

The above formul\ae\ for $K^\pm$ rates do not include particle losses in the 
beam--pipe wall and in the internal tracking system, which were assumed 
sufficiently thin ({\sl e. g.} of a few hundred $\mu m$ of low--$Z$ material, 
such as carbon fibers or Mylar), nor rescattering effects in a nuclear target 
such as $^4$He. We have indeed checked that, due to the shape of the angular 
distribution of the kaons, particle losses are contained (mostly at small 
angles, where $K$--production is negligible, and events would anyhow be hard 
to be fully reconstructed), and momentum losses flat around $\theta = \pi / 2$ 
(where most of the $K^\pm$'s are produced): even for a total thickness of the 
above--mentioned materials of 1 $mm$, kaon momenta do not decrease below 100 
$MeV/c$ and losses do not grow beyond a few percents. Rather, one could 
exploit such a thickness as a low--momentum, thin moderator, to span the 
interesting region just above the charge--exchange threshold at $p_L(K^-) 
\simeq$ 90 $MeV/c$, measurements which would add precious, additional 
constraints on low--energy amplitude analyses$^{13}$.

We have presented the above simplified estimates to show that acceptable rates 
can be achieved, orders of magnitude above those of existing data at about the 
same momentum, {\sl i.e.} to the lowest--energy points of the British--Polish 
Track--Sensitive Target (TST) Collaboration, taken in the mid and late 
seventies at the (too hastily closed down) NIMROD accelerator$^{14}$.

Since losses do not affect $K_L$'s, a detector of the kind sketched above, 
similar in geometry to the one proposed by T. Bressani$^9$ to do 
$K^+$--nucleus scattering and hypernuclear experiments, could be used without 
any problem to study low--energy $K_L \to K_S$ regeneration and 
charge--exchange in gaseous targets, providing essential information for 
this kind of phenomena.

We wish to add that a DA$\Phi$NE detector dedicated to kaon experiments on 
gaseous H$_2$ and D$_2$ can continue its active life, without substantial 
changes, to measure $K^+$--, $K^-$--, and $K^0_L$--interactions on heavier 
gases as well (He, N$_2$, O$_2$, Ne, Ar, Kr, Xe), exploring not 
only the properly nuclear aspects of these interactions, such as nucleon 
swelling in nuclei$^{15}$, but also producing $\pi \Sigma$, $\pi \Lambda$ and 
$\pi \pi \Lambda$ systems at invariant masses below the elastic $\bar K N$ 
threshold in the so--called unphysical region, with statistics substantially 
higher than those now available$^{16}$, due to the $\simeq 4 \pi$ geometry 
allowed by a colliding--beam--machine detector.

\vglue 1.0truecm
\noindent{\bf 3. Impact of DA$\Phi$NE on baryon spectroscopy: the states
$\Lambda$(1405) and $\Sigma$(1385). }
\vglue 0.3truecm

At low momenta, comparable to those of the kaons from DA$\Phi$NE, we have 
data from low--statistics experiments, mostly hydrogen bubble--chamber ones 
on $K^- p$ (and $K^-d$) interactions$^{14,18}$ (dating from the early 
sixties trough the late seventies), plus scant data from $K_L$ interactions 
and $K_S$ regeneration on hydrogen$^{19}$.

The inelastic channels, open at a laboratory energy $\omega = {1\over2} 
M_\phi$ (for $K^\pm$'s the value of $\omega$ at the interaction point has 
to include ionization energy losses as well), are the two--body ones 
$\pi \Lambda$ and $\pi \Sigma$ (in all possible charge states), plus the 
three--body ones $\pi \pi \Lambda$ and (marginally) $\pi \pi \Sigma$ for 
$K^-$ or $K_L$ interacting with nucleons: $K^+$--initiated processes are 
(apart from charge exchange) purely elastic in this energy region.

For interactions in hydrogen, the c.m. energy is limited by momentum 
conservation to the initial one, equal (neglecting energy losses) to $w = 
(m_p^2 + \mu_K^2 + m_p M_\phi)^{1/2}$, or $1442.4\ MeV$ for incident 
$K^\pm$'s and $1443.8\ MeV$ for incident $K_L$'s. As already mentioned, 
energy losses for charged kaons can be exploited (using the inner parts 
of the detector as a moderator) to explore $K^- p$ interactions in a {\sl 
limited} momentum range, down to the charge--exchange threshold at $w = 
1437.2\ MeV$, corresponding to a $K^-$ laboratory momentum of about $90\ 
MeV/c$.

For interactions in nuclei, momentum can be carried away by spectator 
nucleons, and the inelastic channels can be explored down to threshold. The 
possibility of reaching energies below the $\bar K N$ threshold allows 
exploration of the unphysical region, containing two resonances, the $I = 0$, 
$S$--wave $\Lambda(1405)$ and the $I = 1$, $J^P = {3\over2}^+$ $P$--wave 
$\Sigma(1385)$, observed mostly in production experiments (and, in the first 
case, in very limited statistics ones$^{16}$): the information on their 
couplings to the $\bar K N$ channel relies {\sl entirely} on extrapolations 
of the low--energy $\bar K N$ data. The coupling of the $\Sigma(1385)$ to the 
$\bar K N$ channel, for instance, can be determined via forward dispersion 
relations involving the total sum of data collected at $t \simeq 0$, but still 
with uncertainties which are, {\sl at their best}, still of the order of 50 \% 
of the flavour--$SU(3)$ symmetry prediction$^{20}$; as for the 
$\Lambda(1405)$, even its spectroscopic classification is an open problem, 
{\sl vis--\`a--vis} the paucity and (lack of) quality of the best available 
data$^{21}$. We could add that recently even the presence of a second state 
with the same quantum number has been claimed$^{22}$, and to prove (or 
disprove) such a claim would of course be rather important for the role the 
state has both for kaonic atoms and the determination of the low--energy 
parameters of the kaon--nucleon interactions.

A formation experiment on {\sl bound} nucleons, in an (almost) $4 \pi$ 
apparatus with good efficiency and resolution for low--momentum $\gamma$'s 
(such as KLOE$^2$), can measure a channel such as $K^- p \to \pi^0 \Sigma^0$ 
(above threshold), or $K^- d \to \pi^0 \Sigma^0 n_s$ (both above and below 
threshold), which is pure $I = 0$: up to now all analyses on the 
$\Lambda(1405)$ have been limited to charged channels$^{16}$, and assumed the 
$I = 1$ contamination to be either negligible or smooth and non--interfering 
with the resonance signal. Since the models proposed for the $\Lambda(1405)$ 
differ mostly in the details of the resonance shape, rather than in its 
couplings, and it is precisely the shape which could be changed even by a 
moderate interference with an $I = 1$ background, such measurements would be 
decisive. Having in the same apparatus and at almost the same energy {\sl 
tagged} $K^-$ and $K_L$ produced at the same point, one can further separate 
$I = 0$ and $I = 1$ channels with a minimum of systematic uncertainties, by 
measuring all channels $K_L p \to \pi^0 \Sigma^+$, $\pi^+ \Sigma^0$ and $K^- 
p \to \pi^- \Sigma^+$, $\pi^+ \Sigma^-$, besides, of course, the 
above--mentioned, pure $I = 0$, $K^- p \to \pi^0 \Sigma^0$ one. It must be 
noted that the recent claim for {\sl two} $\Lambda$(1405) states$^{22}$ is 
based on a {\sl very low--statistics measurement}$^{23}$ of the reaction $K^- 
p \to \pi^0 \pi^0 \Sigma^0$ (analysed, we incidentally add, {\sl without} 
inclusion of the well--known low--mass enhancement in the $I = J = 0$ $\pi 
\pi$ channel, sometimes known as the $\sigma$--meson!): an analysis of {\sl 
all} $\pi \pi Y$ ($Y = \Lambda, \Sigma$) channels, possible with much higher 
statistics at DA$\Phi$NE, would be therefore highly desirable.

Another class of inelastic processes which are expected to be produced, 
at a much smaller rate, by DA$\Phi$NE's kaons are the radiative capture 
processes $K^- p \to \gamma \Lambda$, $\gamma \Sigma^0$ and $K_L p \to \gamma 
\Sigma^+$ (both in hydrogen and deuterium), and $K^- n \to \gamma \Sigma^-$ 
and $K_L n \to \gamma \Lambda$, $\gamma \Sigma^0$ (only in deuterium). Up to 
now only searches for photons emitted after stops of $K^-$'s in liquid 
hydrogen and deuterium have been performed with some success: the spectra are 
dominated by photons from unreconstructed $\pi^0$ and $\Sigma^0$ 
decays$^{24}$, and separating the signals from this background poses serious 
difficulties, since only the photon line from the $\gamma \Lambda$ final state 
falls just above the endpoint of the photons from $\pi^0$ decays in the $\pi^0 
\Lambda$ final state, while that from $\gamma \Sigma^0$ falls right on top of 
the latter. Indeed these experiments were able to produce only an estimate of 
the respective branching ratios$^{25}$.

The $4 \pi$ geometry possible at DA$\Phi$NE, combined with the 
``transparency'' of a KLOE--like apparatus$^2$, its high efficiency for 
photon detection and its good resolution for spatial reconstruction of the 
events, should make possible (in an H$_2$/D$_2$ experiment) the full 
identification of the final states and therefore the measurement of the 
absolute cross sections for these processes, although in flight and not at 
rest.

Present data$^{24}$ indicate branching ratios around $0.9 \times 
10^{-3}$ for $K^- p \to \gamma \Lambda$ and $1.4 \times 10^{-3}$ for $K^- p 
\to \gamma \Sigma^0$, with errors of the order of 15 \% on both: most 
models$^{26}$ give the first rate larger than the second, with both values 
consistently higher than the observed ones. Only a cloudy--bag--model$^{27}$ 
exhibits the trend appearing (although only at a $2\sigma$--level, and 
therefore waiting for confirmation by better data) from the first experimental 
determinations, but this is the only respect in which it agrees with the data, 
still giving branching ratios larger than observations by a factor two.

Data are also interpretable in terms of $\Lambda(1405)$ electromagnetic 
transition moments$^{25}$: this interpretation is clearly sensitive to the 
interference between the decay of this state and all other contributions. An 
extraction of the $\Lambda(1405)$ moments freer of these uncertainties would 
require measurements of $\gamma \Lambda$ and $\gamma \Sigma$ (if possible, in 
different charge states) over the unphysical region, using (gaseous) deuterium 
or helium as a target. Rates are expected to be of the order of $10^4\ 
events/y$ only, but such a low rate would correspond to better statistics than 
those of the best experiment performed on the $\Lambda(1405) \to \pi \Sigma$ 
decay spectrum$^{16}$.
\vfill
\eject

\vglue 1.0truecm
\noindent{\bf 4. Final recommandations. }
\vglue 0.3truecm

A first, modest proposal would therefore be the following: before building a 
{\sl dedicated apparatus} for low--energy experiments on various gaseous 
targets, one could equip the existing experiment KLOE with a less restrictive 
trigger, that could select the interactions of anti--kaons (tagged by their 
antiparticles on the opposite side, be they either $K^+$'s or $K^0_S$'s) with 
the gas filling the chamber and reconstruct off-line the pion--hyperon, 
pion--pion--hyperon and single--$\gamma$--hyperon spectra for {\sl all} 
charge combinations. Such data would contain both the $\Lambda$(1405) and the 
$\Sigma$(1385), including their interference, plus the effects of 
rescattering {\sl inside} the remainder of the $^4$He target. The latter 
will further feed -- via charge--exchange processes -- also such ``exotic'' 
combinations as $\Sigma^{\pm}\pi^{\pm}$, allowing a better understanding of 
the nuclear--medium distortions on the ``elementary'' processes $\bar K N \to 
\pi Y$, $\bar K N \to \pi \pi Y$ and $\bar K N \to \gamma Y$.

We wish to end underlining how KLOE (or a similar, scaled down apparatus) 
is unique for such a scope: the need for a good efficiency and high 
resolution for low--energy $\gamma$'s (motivated for KLOE by decays such 
as $\phi \to \gamma (a_0,f_0)$ and the reconstruction of very low--momentum 
$\pi^0$'s) allows also the identification and reconstruction of $\Sigma^0$'s 
through their decay to $\Lambda \gamma$, virtually impossible in any other 
detector with an almost 100 \% efficiency. On the other hand, the very high 
efficiency for $\gamma$ detection, combined with the high intensity of the 
source and the ease with which one can discriminate between kaons and pions 
(not to mention leptons) from the $\phi$ decays, allows an unprecedently 
clean determination of radiative capture events (even if in a slightly 
more complex target than hydrogen or deuterium).

As a closing remark one can add that contaminations due to the presence of 
a small admixture of other gases in helium, or to the tungsten wires running 
across the chamber, are not that important for the mass spectra (they amount 
to -- small -- distortions in the nucleon distribution functions, which the 
``elementary'' amplitudes have to be convoluted with, with respect to those 
for pure $^4$He), and even less for the ratio of $\gamma Y$ (or $\pi \pi Y$) 
to $\pi Y$ spectra.

\vfill\eject
\vglue 1.0truecm
\centerline{\bf REFERENCES }
\vglue 0.3truecm

\item{1.} See the Proceedings of the {\sl ``Workshop on Physics and 
Detectors for DA$\Phi$NE''}, G. Pancheri ed. (I.N.F.N., Frascati 1991).

\item{2.} A. Aloisio, {\sl et al.} (The KLOE Collaboration): {\sl ``KLOE 
-- A General Purpose Detector for DA${\mit \Phi}$NE''}, report {\sl 
LNF--92/019 (IR)} (Frascati, April 1992); J. Lee-Franzini: {\sl ``The Second 
DA$\Phi$NE Physics Handbook''}, L. Maiani, G. Pancheri and N. Paver eds. 
(I.N.F.N., Frascati 1995), Vol. II, p. 761.

\item{3.} S. Eidelman, {\sl et al.} (Particle Data Group): {\sl Phys. Lett.} 
{\bf B 592} (2004) 1.

\item{4.} P.M. Gensini and G. Violini: {\sl ``Workshop on Science at the 
KAON Factory''}, D.R. Gill ed. (TRIUMF, Vancouver 1991), Vol. II, p. 193; 
P.M. Gensini: {\sl ``Workshop on Physics and Detectors for DA$\Phi$NE''}, G. 
Pancheri ed. (I.N.F.N., Frascati 1991), p. 453, {\sl ``Common Problems and 
Trends of Modern Physics''}, T. Bressani, S. Marcello and A. Zenoni eds. 
(World Scientific, Singapore 1992), p. 387 and {\sl ``The Second DA$\Phi$NE 
Physics Handbook''}, L. Maiani, G. Pancheri and N. Paver eds. (I.N.F.N., 
Frascati 1995), Vol. II, p. 739; P.M. Gensini, R. Hurtado and G. 
Violini: {\sl $\pi N$ Newslett.} {\bf 13} (1997) 296 and {\sl Genshikaku 
Kenky\u u} {\bf 48}, N. 4 (1998) 51.

\item{5.} C.J. Batty and A. Gal: {\sl Nuovo Cimento} {\bf 102 A} (1989) 
255; C.J. Batty: {\sl ``First Workshop on Intense Hadron Facilities and 
Antiproton Physics''}, T. Bressani, F. Iazzi and G. Pauli ed. (S.I.F., 
Bologna 1990), p. 117. For the latest data, see: KEK experiment: M. Iwasaki, 
{\sl et al.}: {\sl Phys. Rev. Lett.} 
{\bf 78} (1997) 2067; T. Ito, {\sl et al.}: {\sl Phys. Rev.} {\bf C 58} (1998) 
2366. DEAR Collaboration: G. Beer, {\sl et al.}: {\sl Phys. Rev. Lett.} {\bf 
94} (2005) 212302.

\item{6.} J.A. Oller, J. Prades and M. Verbeni: {\sl Phys. Rev. Lett.} {\bf 95} 
(2005) 172502, {\tt arXiv: hep-ph/0508081}. and {\tt ArXiv: hep-ph/0601109}; B. 
Borasoy, R. Ni\ss ler and W. Weise: {\tt arXiv: hep-ph/0512379}.

\item{7.} J. Beveridge: {\sl ``Workshop on Science at the KAON Factory''}, 
D.R. Gill ed. (TRIUMF, Vancouver 1991), Vol. I, p. 19.

\item{8.} G. Vignola: {\sl ``Workshop on Physics and Detectors for 
DA$\Phi$NE''}, G. Pancheri ed. (I.N.F.N., Frascati 1991), p. 11.

\item{9.} T. Bressani: {\sl ``Workshop on Physics and Detectors for 
DA$\Phi$NE''}, G. Pancheri ed. (I.N.F.N., Frascati 1991), p. 475; {\sl 
``Common Problems and Ideas of Modern Physics"}, T. Bressani, B. Minetti and 
A. Zenoni eds. (World Scientific, Singapore 1992), p. 222; {\sl ``Common 
Problems and Trends of Modern Physics''}, T. Bressani, S. Marcello and A. 
Zenoni eds. (World Scientific, Singapore 1992), p. 241.

\item{10.} See the Proceedings of the {\sl ``Workshop on Science at the KAON 
Factory''}, D.R. Gill ed. (TRIUMF, Vancouver 1991).

\item{11.} {\sl ``Proposal for a European Hadron Facility''}, J.F. Crawford 
ed., report {\sl EHF--87--18} (Trieste -- Mainz, may 1987).

\item{12.} J.K. Kim: {\sl Phys. Rev. Lett.} {\bf 19} (1967) 1074, and his 
Ph.D. Thesis, Columbia University Report NEVIS--149 (New York 1966).

\item{13.} See the discussion on this point by D.J. Miller: {\sl ``Proc. of 
the Int. Conf. on Hypernuclear and Kaon Physics''}, B. Povh ed. (M.P.I., 
Heidelberg 1982), p. 215.

\item{14.} TST Collaboration: R.J. Novak, {\sl et al.}: {\sl Nucl Phys.} {\bf 
B 139} (1978) 61; N.H. Bedford, {\sl et al.}: {\sl Nukleonika} {\bf 25} (1980) 
509; M. Goossens, G. Wilquet, J.L. Armstrong and J.H. Bartley: {\sl ``Low and 
Intermediate Energy Kaon--Nucleon Physics''}, E. Ferrari and G. Violini eds. 
(D. Reidel, Dordrecht 1981), p. 131; J. Ciborowski, {\sl et al.}: {\sl J. 
Phys.} {\bf G 8} (1982) 13; D. Evans, {\sl et al.}: {\sl J. Phys.} {\bf G 9} 
(1983) 885; J. Conboy, {\sl et al.}: {\sl J. Phys} {\bf G 12} (1986) 1143. A 
good description of the experiment is in D.J. Miller, R.J. Novak and T. 
Tyminiecka: {\sl ``Low and Intermediate Energy Kaon-Nucleon Physics''}, E. 
Ferrari and G. Violini eds. (D. Reidel, Dordrecht 1981), p. 251.

\item{15.} E. Piasetzky: {\sl Nuovo Cimento} {\bf 102 A} (1989) 281.

\item{16.} R.J. Hemingway: {\sl Nucl. Phys.} {\bf B 253} (1985) 742. Older 
data are even poorer in statistics: see ref. 17 for a comparison. See also, 
for formation on bound nucleons, B. Riley, I.T. Wang, J.G. Fetkovich and J.M. 
McKenzie: {\sl Phys. Rev.} {\bf D 11} (1975) 3065.

\item{17.} R.H. Dalitz and S.F. Tuan: {\sl Ann. Phys. (N.Y.)} {\bf 10} (1960) 
307.

\item{18.} $K^\pm p$ data: W.E. Humphrey and R.R. Ross: {\sl Phys. Rev.} {\bf 
127} (1962) 1; G.S. Abrams and B. Sechi--Zorn: {\sl Phys. Rev.} {\bf 139} 
(1965) B 454; M. Sakitt, {\sl et al.}: {\sl Phys Rev} {\bf 139} (1965) B 719; 
J.K. Kim: Columbia Univ. report {\sl NEVIS--149} (New York 1966) and {\sl 
Phys. Rev. Lett.} {\bf 14} (1966) 615; W. Kittel, G Otter and I. Wa\v{c}ek: 
{\sl Phys. Lett} {\bf 21} (1966) 349; D. Tovee, {\sl et al.}: {\sl Nucl. 
Phys.} {\bf B 33} (1971) 493; T.S. Mast, {\sl et al.}: {\sl Phys. Rev.} {\bf 
D 11} (1975) 3078 and {\bf D 14} (1976) 13; R.O. Bargenter, {\sl et al.}: {\sl 
Phys. Rev.} {\bf D 23} (1981) 1484. $K^- d$ data: R. Armenteros, {\sl et al.}: 
{\sl Nucl. Phys.} {\bf B 18} (1970) 425.

\item{19.} $K_Lp$ data: J.A. Kadyk, {\sl et al.}: {\sl Phys. Lett.} {\bf 17} 
(1966) 599 and report {\sl UCRL--18325} (Berkeley 1968); R.A. Donald, {\sl et 
al.}: {\sl Phys. Lett.} {\bf 22} (1966) 711; G.A. Sayer, {\sl et al.}: {\sl 
Phys. Rev.} {\bf 169} (1968) 1045.

\item{20.} G.C. Oades: {\sl Nuovo Cimento} {\bf 102 A} (1989) 237.

\item{21.} A.D. Martin, B.R. Martin and G.G. Ross: {\sl Phys. Lett.} {\bf B 
35} (1971) 62; P.N. Dobson jr. and R. McElhaney: {\sl Phys. Rev.} {\bf D 6} 
(1972) 3256; G.C. Oades and G. Rasche: {\sl Nuovo Cimento} {\bf 42 A} (1977) 
462; R.H. Dalitz and J.G. McGinley: {\sl ``Low and Intermediate Energy 
Kaon--Nucleon Physics''}, E. Ferrari and G. Violini eds. (D. Reidel, Dordrecht 
1981), p. 381, and the Ph.D. thesis by J.G. McGinley (Oxford Univ. 1979); 
G.C. Oades and G. Rasche: {\sl Phys. Scr.} {\bf 26} (1982) 15; J.P. Liu: {\sl 
Z. Phys.} {\bf C 22} (1984) 171; B.K. Jennings: {\sl Phys. Lett.} {\bf B 176} 
(1986) 229; P.B. Siegel and W. Weise: {\sl Phys. Rev.} {\bf C 38} (1988) 2221; 
R.H. Dalitz and A. Deloff: {\sl J. Phys.} {\bf G 17} (1991) 289.

\item{22.} E. Oset, V.K. Magas, A. Ramos: {\tt 
arXiv: nucl-th/0512090} and {\tt  arXiv: hep-ph/0512361}.

\item{23.} S. Prakhov, {\sl et al.} (Crystall Ball Collaboration): {\sl Phys. 
Rev.} {\bf C 70} (2004) 034605.

\item{24.} B.L. Roberts: {\sl Nucl Phys} {\bf A 479} (1988) 75c; B.L. Roberts, 
{\sl et al.}: {\sl Nuovo Cimento} {\bf 102 A} (1989) 145; D.A. Whitehouse, 
{\sl et al.}: {\sl Phys. Rev. Lett.} {\bf 63} (1989) 1352.

\item{25.} See the review by J. Lowe: {\sl Nuovo Cimento} {\bf 102 A} (1989) 
167.

\item{26.} J.W. Darewich, R. Koniuk and N. Isgur: {\sl Phys. Rev.} {\bf D 
32} (1985) 1765; H. Burkhardt, J. Lowe and A.S. Rosenthal: {\sl Nucl Phys.} 
{\bf A 440} (1985) 653; R.L. Workman and H.W. Fearing: {\sl Phys. Rev.} {\bf D 
37} (1988) 3117; R.A. Williams, C.R. Ji and S. Cotanch: {\sl Phys. Rev.} {\bf 
D 41} (1990) 1449; {\sl Phys. Rev.} {\bf C 43} (1991) 452; H. Burkhardt and J. 
Lowe: {\sl Phys. Rev.} {\bf C 44} (1991) 607. For radiative capture on 
deuterons (and other light nuclei), see: R.L. Workman and H.W. Fearing: {\sl 
Phys. Rev.} {\bf C 41} (1990) 1688; C. Bennhold: {\sl Phys. Rev.} {\bf C 42} 
(1990) 775.

\item{27.} Y.S. Zhong, A.W. Thomas, B.K. Jennings and R.C. Barrett: {\sl 
Phys. Lett.} {\bf B 171} (1986) 471; {\sl Phys. Rev.} {\bf D 38} (1988) 837 
(which corrects a numerical error contained in the previous paper).

\bye